\documentclass[twocolumn]{revtex4}

\usepackage{color}

\usepackage{graphicx}

\def\giorno{19/09/2021 -- revised 29/12/2021}

\begin{document}

\def\a{\alpha}
\def\b{\beta}
\def\ga{\gamma}
\def\de{\delta}
\def\De{\Delta}
\def\la{\lambda}
\def\La{\Lambda}
\def\s{\sigma}
\def\om{\omega}
\def\vphi{\varphi}
\def\eps{\varepsilon}

\def\am{\overline{\a}}

\def\xb{{\bf x}}
\def\yb{{\bf y}}
\def\ub{{\bf u}}
\def\vb{{\bf v}}
\def\wb{{\bf w}}
\def\zb{{\bf z}}
\def\fb{{\bf f}}
\def\pb{{\bf p}}
\def\qb{{\bf q}}
\def\yb{{\bf y}}
\def\Fb{{\bf F}}
\def\Rb{{\bf R}}

\def\xd{\dot{x}}
\def\xdd{\ddot{x}}

\def\zetab{\zeta}

\def\L{\mathcal{L}}
\def\G{\mathcal{G}}

\def\grad{\nabla}

\def\pa{\partial}
\def\d{{\mathrm d}}
\def\T{\mathrm{T}}
\def\E{\mathcal{E}}

\def\^#1{\widehat{#1}}
\def\wh#1{\widehat{#1}}
\def\wt#1{\widetilde{#1}}

\def\({\left(}
\def\){\right)}
\def\[{\left[}
\def\]{\right]}

\def\<{\langle}
\def\>{\rangle}

\def\rosso{\color{red}}
\def\ro{\color{red}}
\def\blu{\color{blue}}
\def\verde{\color{green}}
\def\arancio{\color{orange}}
\def\viola{\color{violet}}

\def\gcite#1{{\blu \cite{#1}}}

\def\beq{\begin{equation}}
\def\eeq{\end{equation}}

\def\EOR{\hfill $\odot$}

\def\beql#1{\begin{equation} \label{#1}}

\def\eqref#1{(\ref{#1})}

\def\wm{\< w \>}
\def\am{\< \a \>}
\def\P{\mathcal{P}}





\title{Mass vaccination in a roaring pandemic}
\label{firstpage}

\author{Giuseppe Gaeta}

\affiliation{{\it Dipartimento di Matematica, Universit\`a degli Studi di Milano,  v. Saldini 50, I-20133 Milano (Italy)} \\
\& \  {\it SMRI, 00058 Santa Marinella (Italy) }\\ {\tt giuseppe.gaeta@unimi.it}}

\date{\giorno}

\begin{abstract}

Mass vaccination produces a reduction in virus circulation, but also evolutive pressure towards the appearance of virus-resistant strains. We discuss the balance between these two effects, in particular when the mass vaccination takes place in the middle of an epidemic period.

\end{abstract}

\maketitle





\section{Introduction}

\noindent
One of the basic -- but rather counter-intuitive -- results in Mathematical Evolution Theory is the \emph{principle of competitive exclusion} \gcite{Gause,HS,Murray}. This says that if two variants of an organism, having slightly different fitness, compete for the same resources (e.g. two strains of a virus compete to infect the same population), then the one with the higher fitness will completely eliminate the other, albeit not instantaneously (the time needed for this will be a decreasing function of the fitness difference).

This is counter-intuitive, as one could think there would be an equilibrium with the fittest one representing the highest part of the population; but it is essentially a reformulation of Darwin's \emph{survival of the fittest}.

This principle raises a question related to mass vaccination campaigns taking place when there is a high viral circulation, e.g. in the middle of a roaring pandemic as it happens in present days for COVID.

That is, on the one hand the vaccination will reduce the amount of virus circulating in the population, and so decrease the probability of any individual to be infected by any strain of the virus; but on the other hand, the presence of a large amount of vaccinees makes that the fitness of variants which are able to (totally or partially) elude the vaccine -- more precisely, the antibodies stimulated by the vaccine -- is higher than that of strains which are effectively countered by the vaccine. So, \emph{if} a vaccine-resistent strain arise by random mutations, \emph{then} it will have an easy life and will promptly discard the previously dominant strain on which the vaccine was effective).

This is not just a theoretical picture: e.g. we know that this is precisely what happened (in the sense of a strain with higher fitness replacing the wild type; luckily in this case the new strain has only a marginal resistance to existing vaccines) when the so called \emph{alpha variant} replaced the original strain of COVID-19. Actually, it is quite surprising that this replacement took place following extremely well \gcite{Fort} the theoretical predictions of the most basic models \gcite{Gause,HS,Murray}.

The difference in fitness among different variants will depend in part on the molecular biology of the virus, but also in part on the environment and on the measures taken to counter the epidemic. While physical and chemical barriers (face masks, UV light, frequent use of disinfectants) are expected to act more and less in the same way on different strains, the situation is quite different for the vaccine action -- in particular for ``modern type'' vaccines as those used for the COVID, and targeting very specifically a given strain (the one existing at the time the vaccine is studied and produced). In this case, mutations can lead to variants which can elude the action of the vaccine.

The situation is even more extreme in the COVID case: in this case, indeed, the vaccine targets specifically the \emph{spike} protein, which is essential in binding to cell receptors, and not the core of the virus itself, which enters in action once the virion penetrates the cell. A mutation in the spike can bypass (with more or less high probability) the specific antibodies induced by the vaccine, and if this happens the virions which manage to enter the cell have no special antibodies originating in the vaccine to contrast its replication.\footnote{We recall that, luckily, the immune system manages to react quite effectively to the COVID infection. E.g., estimations based on the early phase of the pandemic suggested that the infection is symptomatic only in a minority of cases, estimations ranging form one out of seven \gcite{Li} to one out of ten \gcite{CG,Gaeta}. Asymptomatic cases are of course as bad as symptomatic ones -- or even worse, given that the carriers can escape detection and thus circulate freely in the population -- for what concerns infection dynamics, but are less dangerous for what concerns consequences on the carrier; and one should not forget that they represent cases in which the immune system managed to cope perfectly with the virus.}

The dilemma takes a different aspect when -- again, as in the COVID case -- a large part of the population, e.g. young and healthy ones, typically has very little consequences due to the infection; while another part of the population, e.g. old and fragile ones, is quite vulnerable and at risk of death, or anyway of serious consequences, in case of infection. Giving for granted that this latter part of the population should be vaccinated as soon and widely as possible, it is not obvious that a mass vaccination of the former part of the population does not, in the long run, \emph{increase} the risk of having variants which can elude the vaccine and thus put again at risk the older part of the population, which was screened from infection by the first run of vaccinations.\footnote{One should recall, in this context, that the recent work by Fokas, Cuevas-Maraver and Kevrekidis \gcite{Fokas2} showed that extending lockdown measures to young and healthy people does not really reduce the number of expected casualties; we will come back to this point in Sect.\ref{sec:discussion}.}

Thus, in summary, on the one hand mass vaccination reduces the circulation of the virus and thus the occasion of mutations, including the dangerous ones, while on  the other hand it increases the probability that if a dangerous mutation occurs it will quickly spread among the population.\footnote{In the case of COVID, the situation is made even more complex by the fact that while in the developed countries most of the population (above some age) is vaccinated, this is not the case -- and it is actually impossible for logistic reasons (e.g. vaccines needing very low temperatures) -- in developing or even more in poor countries. Thus a large part of the world is not touched by the vaccination campaign, and can act as a source of variants with a high virus circulation.}

Obviously there is no way to evaluate the net result of these two contrasting effects in qualitative terms: one should produce some quantitative discussion, possibly based on simplified models, to understand the balance between these two contrasting effects. This is precisely what we are doing in this note.\footnote{A number of side remarks or personal views will be given in footnotes.}

\section{Mutation vs vaccination}

The mutation rate of the virus is, as far as we know, not affected by vaccination. We assume each amino-acid in the virus sequence\footnote{Note we are thus working at the phenotype level, not at the genotype one. This is appropriate as public data about COVID mutations are given in this form \gcite{Stanford}.} can mutate with a probability $\mu_0$ in a replication (this represents an average for substitution, insertion, or deletion mutations). Thus the probability of a given mutation appearing within a time $\tau$ from some arbitrarily chosen ``initial time'' depends on the phenotypical distance $d$ between the original strain and the mutated one (there should be $d$ mutations in the given sites and no other ones) \emph{and} on the amount of circulating virus, which is directly proportional to the amount of replications occurring in viral matter. We are mostly interested in this second aspect of the matter.

We assume that vaccination is effective to a rate $\gamma$, i.e. that out of $n$ vaccinees getting in contact with the virus, only $(1-\gamma) n$ will have a chance to contract the infection, while $\gamma n$ will be fully shielded from it.

Thus the number of replications in the total viral population will also be reduced by a factor $(1 - \gamma)$, and so will be the probability $\P (V;d,\ell;\tau)$ that the variant $V$ with phenotypical distance $d$ (out of a phenotype of length $\ell$) will appear within time $\tau$.

More precisely, we will denote by $\P_{\chi,\gamma} (V;d,\ell;\tau)$ this probability in a population with a fraction $\chi \in [0,1]$ of the population being vaccinated with a vaccine having effectiveness $\gamma$, and $\P_0 (V;d,\ell;\tau)$ the same probability in a population where nobody is vaccinated (this is independent of $\ga$). Our previous discussion shows then that
$$ P_{1,\gamma} (V;d,\ell;\tau) \ \approx \ (1 - \gamma ) \ \P_0 (V;d,\ell;\tau) \ . $$
By the same line of reasoning, considering a partially vaccinated population (and omitting $V;d,\ell$ for ease of notation, these being fixed) we get
\begin{eqnarray}
P_{\chi,\gamma} (\tau) & \approx & \chi \, P_{1,\gamma} (\tau) \ + \ (1 - \chi ) \, P_0 (\tau)  \nonumber \\
& \approx & \[ \chi \, (1 - \gamma ) \ + \ ( 1 - \chi ) \] \ P_0 (\tau) \label{eq:P0tau} \\
&=& (1 \ - \ \chi \, \gamma ) \ P_0 (\tau) \ . \nonumber \end{eqnarray}

Equivalently, the expected time $\< \tau \>_{\chi,\gamma}$ for the appearance of the variant in a population with a fraction $\chi$ of vaccinees and vaccine effectiveness $\gamma$ relates to the expected time $\< \tau \>_0$ in the same fully non vaccinated population via
\beql{eq:P0taubis} \< \tau \>_{\chi,\gamma} \ \approx \ \( \frac{1}{1 \ - \ \chi \, \gamma } \) \ \< \tau \>_0 \ . \eeq

Estimating $\P_0 (\tau) = \P_0 (v;d,\ell;\tau)$ is not of direct interest here, as we are interested in how this probability changes depending on the level of vaccination in the population and on the effectiveness of vaccine, i.e. on \eqref{eq:P0tau}.

It should be stressed that when we speak of vaccine effectiveness, we do \emph{not} mean this in medical sense, but in epidemiological one. That is, we are not considering what is the protection offered by vaccine against serious consequences from infection, but the bare protection from getting infected and infective; see the brief discussion in Appendix \ref{app:asprot}.

\section{Fitness versus vaccination}

The fitness of a variant in a fully non vaccinated population is essentially a random variable; we know that only variants with fitness higher than the dominant type will appear in a sizeable fraction of the population, so fitness of the dominant type will be an increasing function of time if measured in a ``virus virgin'' population.

On the other hand, if the population has already been in contact with the virus, there is acquired immunity against the strain each individual has been in contact with, and also against strains which are genetically near enough to these. How near is a matter which depends on the virus at hand (and also, to a lesser degree, on the individual immune system); but one can safely assume, based on experience with other viruses, that the variants appearing in a few years -- at least under a ``natural'' evolution -- will be near enough to be recognized by the immune system of most people (or at least of most people having survived the infection).

In this scenario, the ``susceptible'' population\footnote{Of course the name ``susceptible'' reminds of the familiar SIR (susceptible / infected and infective / removed) model for contagious diseases, and its variants.} is decreasing in time, and thus also the fitness of any given variant. In other words, albeit the natural evolution leads to strains with higher fitness and thus higher infectivity, the accumulation of antibodies makes that the fitness of any given strain in the same population decreases in time. In the long run this will lead to a low fitness, but one can have a transient behavior with temporarily high fitness and thus infectivity.

All this discussion, however, does not take into account the change in environment due to the introduction of vaccines.

Consider a strain of fitness $\phi_0$ in a fully non-vaccinated population, and of fitness $\phi_1$ in a fully vaccinated population. In a partially vaccinated  population, with a fraction $\chi$ of vaccinees, its fitness will be
\beq \phi_\chi \ = \ (1 - \chi) \, \phi_0 \ + \ \chi \, \phi_1 \ . \eeq
It is notationally convenient to assume that
\beq \phi_1 \ = \ (1 - \ga ) \ \phi_0 \ , \eeq
with $\ga \in [0,1]$ representing the vaccine efficiency against virus reproduction. Then we get
\beql{eq:phichi} \phi (\chi) \ = \ \( 1 \, - \, \ga \, \chi \) \ \phi_0 \ . \eeq

This is the same result obtain in the previous Section, see \eqref{eq:P0tau}, under a (slightly) different description.

\section{Fitness versus lethality}

We have so far discussed quantities related to the speed of diffusion of (different strains of) a virus. But this in itself is not necessarily interesting: in fact, nobody is (or should be) really worried if a virus diffuses very fast but does not cause any harm. For example, in Italy out of a population of about $6 * 10^7$, in pre-covid (and lockdown) years there were typically $5*10^6$ cases of influenza per year \gcite{iss}, and about $5*10^3$ led (due to complications and concurrent pathologies) to death\footnote{At present (early September 2021), the casualty rate for COVID in Italy is about 1/100 of known infected people. This ratio, however, can not be directly compared with that (1/1000) for influenza: the COVID virus is tracked, and among known infections there are many which are asymptomatic, while for influenza nobody is taking care of, or tracing, asymptomatic or paucisymptomatic infections.}. Despite the huge number of infected people, such a virus is dangerous only for older and fragile people, and it is usually contrasted with a specifically targeted vaccination campaign and through obvious care in contact with people at risk in the epidemic season.

It is well known, and also easily understood, that in general when we compare different strains of the same virus, the more lethal are also less easily transmitted; if not for other reason, because the carriers die and because the contacts of the infected people keep carefully the distance from them in the case of a dangerous virus.

This means that, once there is a widespread awareness\footnote{For an attempt to include awareness -- and consequences of this -- in epidemic modelling, see \gcite{LS}.} about the presence and danger of a virus, there is also a natural selective pressure towards less lethal strains. We are actually witnessing this also with the COVID pandemic: once the general population has realized that the circulating strains (such as the delta variant) are more infective but also less dangerous\footnote{It should actually be recalled that the smaller hospitalization and mortality rates depend also on the fact medical science has found better ways to contrast the infection. In this respect, one could recall that the first guidelines by WHO suggested the use of paracetamol, which was then found to be not effective (or even deleterious) in the treatment of COVID, while general use anti-inflammatory agents such as acetylsalicylic acid proved much better for treatment of patients in their initial stage.} than the original ones, many distancing measures and habits are being relaxed.\footnote{This relaxation is often condemned as dangerous, but it is actually a form of ``popular wisdom'', as it happened over the years with influenza after the catastrophe of the Spanish flu in 1917 and following years.}

We will define lethality, which we denote by $\la$, as the fraction of infected individuals which dies due to the infection; obviously, by definition, $\la \in [0,1]$.

Having established that we generally expect an inverse relation -- when we do not consider vaccination effects -- between fitness $\phi$ and lethality, i.e. that $\phi$ should be a \emph{decreasing} function of $\la$, we prefer not to assume any specific such dependence: this would be arbitrary, and our main result is independent of such a detailed relation. See however Appendix C.

\section{The effects of mass vaccination}

We should now discuss how the results of our simple discussion help us in answering our initial question, i.e. what is the net effect of a mass vaccination campaign conducted under widespread virus circulation.

We will consider a ``present-time'' strain $P$ (denoted by the index $i=1$) with fitness $\phi_1$ in a fully non-vaccinated host population; and a variant $V$ (denoted by the index $i=2$) with fitness $\phi_2$ in a fully non-vaccinated host population; we will consider the cases $\phi_2 > \phi_1$ and $\phi_2 < \phi_1$.

Note that in view of our basic hypothesis of negative correlation between fitness and lethality, this means in these two cases we also have -- with obvious notation -- respectively $\la_2 < \la_1$ and $\la_2 > \la_1$.

The interesting case is the one where the variant $V$ is vaccine-resistant; or at least the vaccine effectiveness against $V$ is substantially smaller than against $P$. In order to consider this case in more general terms, we will consider vaccine effectiveness $\ga_i$, with
\beql{eq:gamma} 0 \le \ga_2 < \ga_1 \le 1 \ . \eeq

We will also denote the effective fitness of the two strains in a partially vaccinated population (with a fraction $\chi$ of vaccinees) as $\phi_i (\chi)$. Thus $\phi_i \equiv \phi_i (0)$.

\subsection{Case A: $\phi_2 > \phi_1$, $\la_2 < \la_1$}

If $\phi_2 > \phi_1$, it is immediately seen that \eqref{eq:gamma} guarantees to have
\beq \phi_2 (\chi ) \ > \ \phi_1 (\chi ) \eeq
for any $\chi \in [0,1]$.

This means that when the variant $V$ (having higher ``naked'' fitness) appears, whatever the level of vaccination in the host population at that time, it will have higher fitness and thus will become prevalent. On the other hand, raising the fraction of vaccinees -- i.e. raising $\chi$ -- will lead to a delay in the appearance of the variant $V$, as seen above in eq. \eqref{eq:P0tau} and \eqref{eq:P0taubis}.

Note that in this case we are talking about a \emph{less lethal} variant, so a delay in its appearance is actually an unwanted feature.

\subsection{Case B: $\phi_2 < \phi_1$, $\la_2 > \la_1$}
\label{sec:VB}

Let us now consider the case where $\phi_2 < \phi_1$, and hence $\la_2 > \la_1$. This is a more worrying case, as the variant is more lethal; in the absence of vaccination this has a lower fitness so even if it is produced by random mutation, it will not survive in the population.

On the other hand, as we supposed it to be vaccine resistant (at least in the sense of being less affected by the available vaccines), now the relation $\phi_2 < \phi_1$ is \emph{not} guaranteed to be preserved when we consider a nonzero $\chi$. In this case \eqref{eq:phichi} yields, with obvious notation,
\beql{eq:phichiV} \phi_i (\chi ) \ = \ \( 1 \ - \ \ga_i \, \chi \) \ \phi_i (0) \ . \eeq

Then the inequality $\phi_2 (\chi) < \phi_1 (\chi )$ transforms, under successive trivial operations, into
\begin{eqnarray*}
(1 - \ga_2 \chi ) \, \phi_2 &<& (1 - \ga_1 \chi) \, \phi_1 \ , \\
(\phi_2 - \phi_1 ) &<& (\ga_2 \phi_2 - \ga_1 \phi_1 ) \, \chi \ , \\
\phi_1 - \phi_2 &>& (\ga_1 \phi_1 - \ga_2 \phi_2 ) \, \chi \ . \end{eqnarray*}
In other words, the fitness of the (more lethal) variant $V$ in a partially vaccinated population is less than that of the original strain $P$ \emph{only} if the fraction of vaccinees in the host population is smaller than a certain critical value $\chi_*$, and more precisely if it satisfies
\beql{eq:chicrit} \chi \ < \ \chi_* \ := \ \frac{\phi_1 \ - \ \phi_2}{\ga_1 \, \phi_1 \ - \ \ga_2 \, \phi_2 } \ . \eeq

This relation is more conveniently expressed in terms of the ratios
\beq p \ := \ \frac{\phi_2}{\phi_1} \ , \ \ q \ := \ \frac{\ga_2}{\ga_1} \ , \eeq
where of course $p,q$ are positive parameters in the interval $(0,1)$. Then we get
\beql{eq:chipq} \chi_* \ = \ f(p,q;\ga_1) \ := \ \frac{1 \ - \ p}{\ga_1 \ (1 \, - \, p \, q )} \ ; \eeq
or more precisely, recalling that $0 < \chi_* \le 1$ and considering $\ga_1$ as a given parameter, we have
\beql{eq:chipqF} \chi_* \ = \ F(p,q) \ = \ \mathtt{min} \{ f(p,q;\ga_1) \, , \, 1 \} \ . \eeq
The function $\chi_* = F(p,q)$ given by \eqref{eq:chipqF} is plotted in Fig.\ref{fig:a3D} and in Fig.\ref{fig:acont} for different values of $\ga_1$, i.e. for $\ga_1 = 1$ (this is the case the vaccine is 100\% effective against strain P), for $\ga_1 = 0.9$, and for $\ga_1 = 0.75$.

In order to understand the behavior of $\chi_*$ as a function of the three parameters $\{ p, q ; \ga_1\}$ it may be worth noting that
\begin{eqnarray*}
\frac{\pa f}{\pa p} &=& - \, \frac{1}{\ga_1} \ \frac{(1-q)}{(1 \, - \, p\, q )^2} \ < 0 \ , \\
\frac{\pa f}{\pa q} &=& \frac{1}{\ga_1} \ \frac{(1-p) \, p}{(1 \, - \, p\, q )^2} \ > 0 \ ; \\
\frac{\pa f}{\pa \ga_1} &=& - \, \frac{1}{\ga_1^2} \ \frac{(1-p)}{1 \, - \, p\, q} \ < 0 \ . \end{eqnarray*}
Thus, the critical vaccination level $\chi_*$ (if different from one) decreases with increasing fitness of the variant $V$ compared with ``present day'' strain $P$, increases with increasing effectiveness of the vaccine against variant $V$ compared with effectiveness against strain $P$, and decreases with increasing effectiveness of the vaccine against the strain $P$.

\begin{figure}
  \includegraphics[width=200pt]{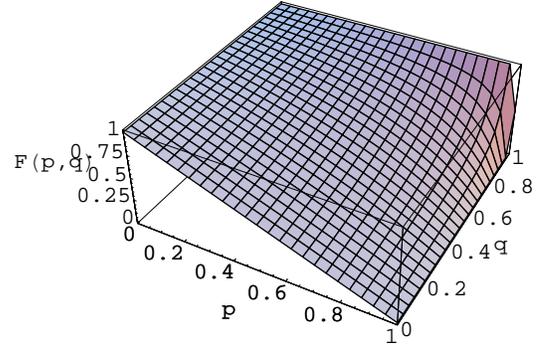}\\
  \includegraphics[width=200pt]{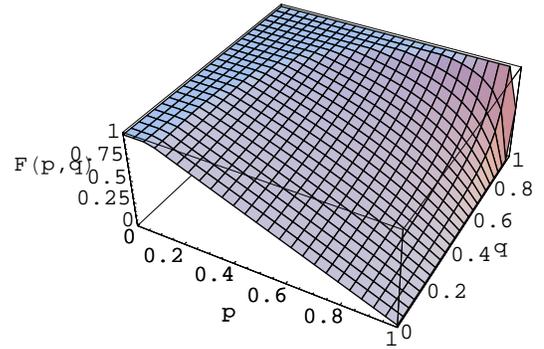}\\
  \includegraphics[width=200pt]{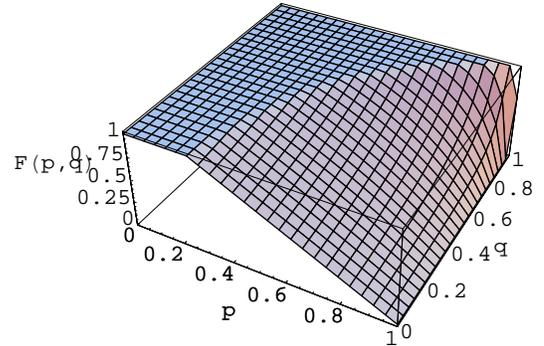}\\
  \caption{Three-dimensional plots of the function $\chi_*  = F (p,q)$ defined in eq.\eqref{eq:chipqF}. We plot the cases $\ga_1 = 1$ (upper), $\ga_1 = 0.9$ (middle), and $\ga_1 = 0.75$ (lower).}\label{fig:a3D}
\end{figure}

\begin{figure}
  \includegraphics[width=150pt]{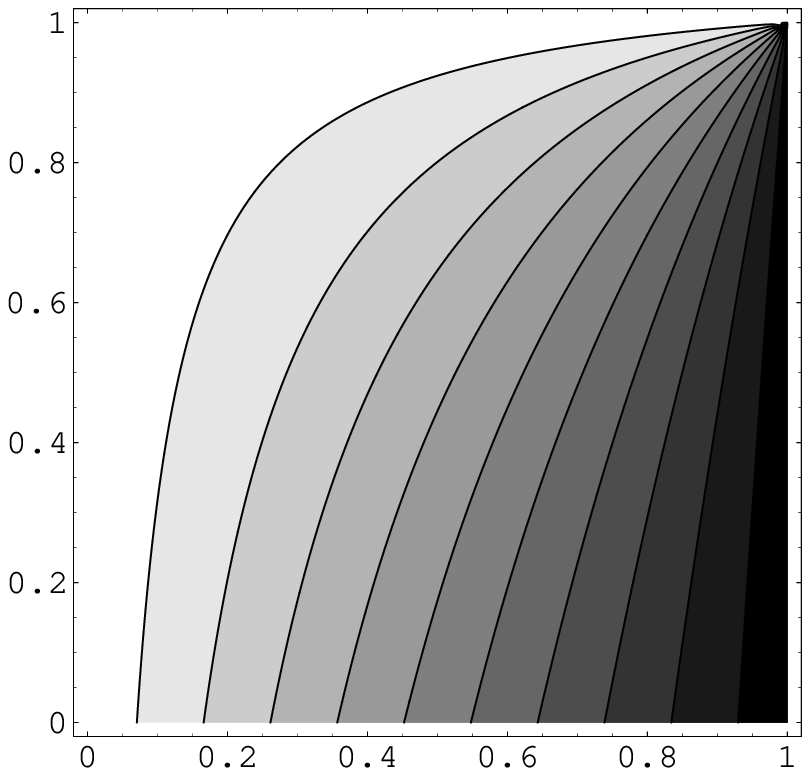}\\
  \includegraphics[width=150pt]{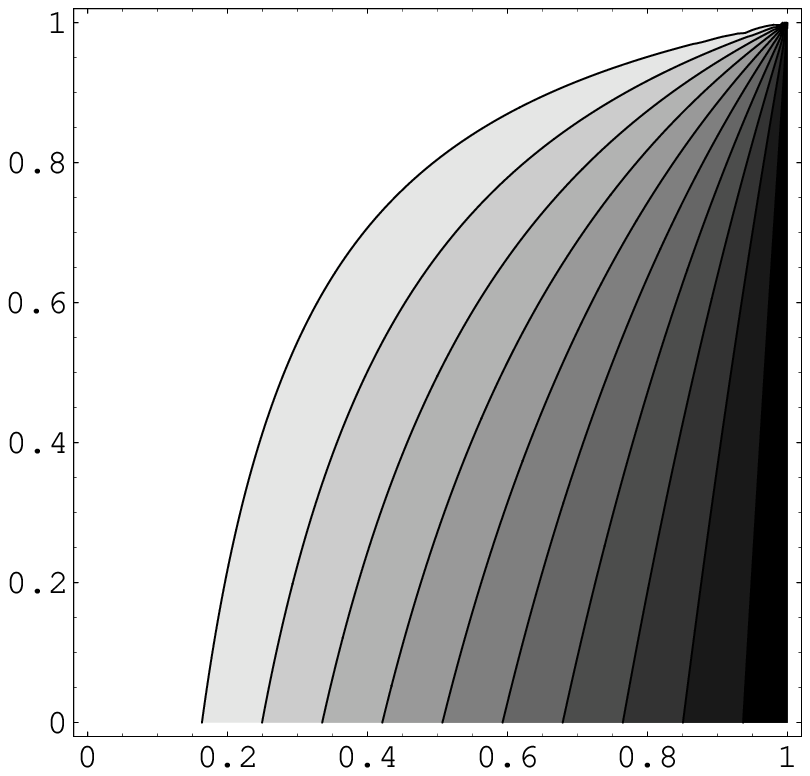}\\
  \includegraphics[width=150pt]{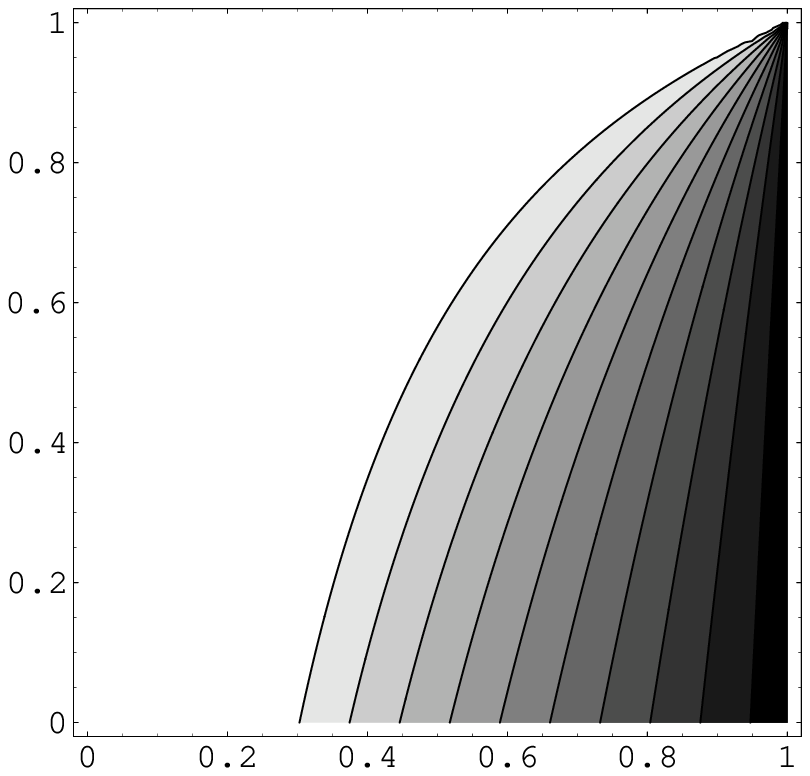}\\
  \caption{Contour plots of the function $\chi_* {\blu = F} (p,q)$ defined in eq.\eqref{eq:chipqF}. We plot the cases $\ga_1 = 1$ (upper), $\ga_1 = 0.9$ (middle), and $\ga_1 = 0.75$ (lower).}\label{fig:acont}
\end{figure}

It may be worth considering the limit case in which the vaccine is completely ineffective against the variant $V$, i.e. $\ga_2 = 0$; this corresponds to $q=0$. In this case we have
$$ f(p,0) \ = \ \frac{1 - p}{\ga_1} \ ; $$
it follows immediately that $\chi_* = 1$ for all $p \le 1 - \ga_1$, but $\chi_* < 1 $ for $1 - \ga_1  < p \le 1$.

We can also consider the ``symmetric'' case in which the reduction in vaccine effectiveness is exactly the same as the reduction in fitness, so $p = q = x$.
Then we have
$$ f(x,x) \ = \ \frac{1-x}{\ga_1 \ (1 - x^2)} \ = \ \frac{1}{\ga_1 \ (1+x) } \ , $$ and the corresponding plot for $\chi_*$ is shown in Fig.\ref{fig:sym}. In this case we have a phase transition at the cusp in
\beql{eq:xstar} x \ = \ x_* \ := \ \frac{1 \, - \, \ga_1}{\ga_1}  \ , \eeq
for $\ga_1 \ge 1/2$.

\begin{figure}
  \includegraphics[width=150pt]{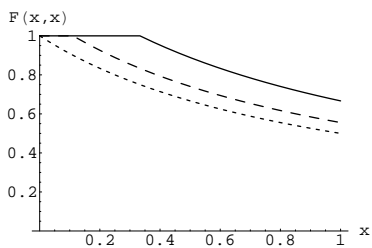}\\
  \caption{The critical vaccination level in the symmetric case $p=q=x$, for $\ga_1 = 0.75$ (solid curve), $\ga_1 = 0.9$ (dashed), and $\ga_1 = 1$ (dotted). The cusp corresponds to $x = x_*$, see text and eq.\eqref{eq:xstar}.}\label{fig:sym}
\end{figure}

Similar results would be obtained, as clear from Fig.\ref{fig:a3D} and Fig.\ref{fig:acont}, for any one-dimensional paths in the $(p,q)$ plane.

These cusps point out a first order phase transition; the critical curve corresponds to the locus of points at which $f(p,q;\ga_1 )= 1$. This can be expressed, see \eqref{eq:chipq}, for any parameter in terms of the other two; in particular we have (denoting by a ``*'' the critical values)
\begin{eqnarray}
p_* &=& \frac{1 \, - \, \ga_1}{1 \, - \, q \, \ga_1} \ , \nonumber \\
q_* &=& \frac{p \, + \, \ga_1 \, - \, 1}{p \, \ga_1} \ . \end{eqnarray}

\subsection{Successive variants and native strain}

In the previous part of this Section, we have compared the fitness of two variants, none of which is the original strain.

We can readily reformulate the intermediate results of that discussion by referring each fitness $\phi_i$  to the fitness of the native strain $N$ (which in the following will be denoted by the index $0$) $\phi_0$ and the vaccine effectiveness $\ga_i$ to that against the native strain $\ga_0$ via
\begin{eqnarray*}
\phi_i \ = \ \kappa_i \ \phi_0 &,& \ \kappa_i > 1 \ , \\
\ga_i \ = \ \s_i \, \ga_0 &,& \ \s_i < 1 \ . \end{eqnarray*}
Then eq.\eqref{eq:chicrit} reads
\beql{eq:chicrit0} \chi \ < \ \chi_* \ := \ \frac{\kappa_1 \ - \ \kappa_2}{(\s_1 \, \kappa_1 \ - \ \s_2 \, \kappa_2) \ \ga_0 } \ . \eeq
On the other hand, eq.\eqref{eq:chipq}, which refers to the ratios $(p,q)$ of fitness and vaccine effectiveness for the two variants, goes unchanged. (Note that with the notation just introduced, we have $p = \kappa_2/ \kappa_1$, $q = \s_2/ \s_1$.)

This formulation has however the advantage of recalling that there is a ``reference'' native strain $N$. The relevance of this is not just ``historical'', or to keep track of the accumulation of mutations; in fact, the vaccine was designed specifically to counter \emph{this} viral strain.\footnote{The design of the vaccine was of course subject to strict time constraint, given the situation the world was in at the time; but in any case one can not try to build a vaccine acting against a too wide class of spike proteins. In fact, the cell receptors -- to which the COVID spike adapts so well -- do have a functional role, and ``good'' biological units must be able to bind to these, and cannot be banned from this. Actually, we are not even completely sure that there are no such ``good'' units having evolved a spike too similar to that of COVID; moreover, by banning such spikes we are renouncing to use these in the future e.g. to deliver specifically targeted (e.g. anti-cancer) drugs within the cell.}

As the vaccine was formulated on the basis of the native strain $N$, its effectiveness against the variants $P$ and $V$ will depend on the differences between $P$ and $N$ and between $V$ and $N$. We know this will \emph{not} be a simple relation: e.g., in the case of COVID, the vaccine is built targeting a specific part of the virus, i.e. the spike -- so that its effectiveness will depend only on mutations occurring in the spike. On the other hand, the action of the virus once it has entered the host cells does \emph{not} depend on the spike, but only on the other, ``core'', part   (note that the lethality of the virus will anyway also depend on its effectiveness to penetrate the cells, i.e. in the case of vaccinated hosts to elude the vaccine action; so we cannot state that the lethality only depends on the ``core'' part of the virus).

A random mutation may raise or (more frequently) lower the virus fitness. When we take into account also the vaccine action, we expect that mutations changing the shape of the spike will more likely be able to elude the vaccine action. Note that these mutation may be harming the functioning of the virus, or at least of the spike, in different ways, or even be fully dysfunctional. What may be said is that \emph{if} a mutation is viable, the fact it make the spike more different from the shape which was targeted when designing the vaccine also means it is more likely to raise its fitness in the environment of a vaccinated host organism.

We know that the natural evolution of a virus is towards strains which have higher and higher fitness and lower and lower lethality. Thus, if we follow the evolution in time of ``present time'' variant $P$, we will have higher and higher $\phi_1 > \phi_0$ (and lower and lower $\la_1 < \la_0$). On the other hand, a vaccine formulated for the native strain $N$ -- and having a very high effectiveness rate $\ga_0$ -- will have smaller and smaller effectiveness against variants which are unavoidably more and more different from $N$.

One should also recall that in the same way as the $\phi_i$ refers to a ``pre-vaccine'' (or ``vaccine-naive'', as is also often called) host population, the same holds for the $\la_i$. In general, we expect that even when the vaccine is not effective in avoiding infection, it will reduce the lethality of the virus\footnote{In the case of COVID, as the vaccine limits the binding of virions to cells, a likely reason for this is that vaccine would reduce the \emph{viral charge} which the immune system has to counter. This consideration resonates with early models of COVID epidemic \gcite{Volpert} aiming at an explanation of the burst-like behavior of the epidemic.}. This effect will again depend on how different the variant at hand is w.r.t. the one on which the virus was gauged, and we expect that on the average successive variants will be less effectively countered by the vaccine also in terms of lethality.

In other words, considering this effect will \emph{enhance} the behavior discussed in the previous subsections.

\section{Discussion, and real-world setting}
\label{sec:discussion}


It should be stressed that we do \emph{not} know which mutations would be ``soft'' and which one would be ``hard'' in terms of ``bare'' (that is, independent of vaccination) infectivity and lethality. On the other hand, as a rule of thumb we can expect that variants with more mutations (in the case of COVID, due to the peculiar nature of vaccines actually used, with more mutations in the spike) would more easily evade the vaccine action.

If taken literally, the results obtained above seems to suggest that vaccination should not be used at all. This is of course a paradoxical conclusion, and one which is in sheer contrast with our consolidated (albeit so far, short-time) experience.

The obvious reason behind this apparent paradox is that the host population is not homogeneous. In the case of COVID, we know that the population of senior and more generally fragile citizens can have very serious consequences from infection, while the population of young and healthy will (unfortunately, on the average and not in all cases) have only mild problems\footnote{The correspondence is not so simple, and most probably some genetic factor is also at play; this makes that some old people have no problems even if infected with COVID, and some young people can have very serious consequences.}, if any at all, from the infection.

It is thus clear that the fragile population should be protected with all the available means; which in the COVID case and at present means distancing measures, individual protection devices, \emph{and vaccines}.

Thus the discussion can only concern the level of vaccination which should be pursued in the population of young and healthy individuals.

In this context, it is appropriate to mention that A.S. Fokas, J. Cuevas-Maraver \& P.G. Kevrekidis \gcite{Fokas2} discussed (before vaccines became available) how a set of differentiated distancing measures for the population of young people and for that of old people could reach the same results than a generalized lockdown in terms of reduction of casualties, easing life of a large part of the population; see also \gcite{Fokas1}.

Our discussion here is in a way an evolutionary counterpart of their one. The net results of this is that once one considers the vaccination of senior and fragile people (and of other parts of the population specially at risk) as granted, then -- quite contrary to the common wisdom -- it may be convenient \emph{in epidemiological terms}   to avoid a generalized vaccination campaign.

We stressed ``in epidemiological terms'' since there are various opinions that this should be avoided for younger people also in medical ones, as the risks connected to vaccination are rather clear and significant, while the infection consequences in this class of the population appear to be (at least of the time of writing) statistically insignificant.\footnote{A study on COVID impact on children \gcite{Smith} estimates the risk of COVID-related death in this age class as 2 in a million; a recent {\it British Medical Journal} editorial discussing COVID vaccination in children \gcite{Saxena} states that ``there are no plans to vaccinate children under 12 in the UK as there is currently insufficient evidence that vaccinating this age group is safe, effective, or acceptable to parents''. See also, in this respect, ref.\gcite{ZC}.}

Moreover, on a long-time vision, it may be argued that exposing young people to the virus infection will provide them with long-time immune memory and thus enable them to live in a world in which the virus has become endemic -- exactly as for influenza\footnote{It may be worth recalling, in this respect, that the only virus which has been eradicated by vaccines is the variole one, and it required a very long term action. Moreover, many poor countries do not have the means to conduct a vaccination campaign. So, all in all and despite the ``war to the virus'' rhetoric, we should expect that -- as for any other virus -- what is realistically attainable is coexistence with a less lethal strain of the virus, and a smaller virus circulation; vaccination is of course a key factor in obtaining the latter.}. This, rather natural, view is also confirmed by the first analysis conducted to compare protection in vaccinees and in patients who have recovered from a COVID infection in Israel \gcite{Gazit} (see also \gcite{Koj,Kojima}).

\section{Conclusions}
\label{sec:conclusions}

We have argued that while enforcing a widespread mass vaccination campaign in the middle of a pandemic will reduce the circulation of the virus, on the other hand it will create a selective advantage for variants able to elude the effect of the vaccine. The net balance between these two contrasting effects is definitely worth investigating, also in view of the ongoing COVID pandemic.

Also, while the risk/benefit balance is definitely in favor of vaccination for senior and fragile citizens, the situation for younger ones is at least unclear, if not definitely with a risk prevalence. This poses an ethical question, i.e. if people which are statistically more prone to have a damage than a benefit from a medical action -- in this case  vaccination -- should be forced to do so for the well being of the whole population. (I am of course not competent to discuss this ethical issue.)

On the epidemiological side, our discussion suggests that a widespread vaccination can lead to: $(a)$ a delay in the appearance of less lethal variants; and $(b)$ allow more lethal variants to become stable in the host population.

The conclusion to be drawn from this is that while vaccination of senior and fragile citizens should definitely be pursued, the situation should be carefully considered before undergoing a generalized vaccination campaign among younger generations, as this could result in a drawback both in the sense of slowing the evolution towards less dangerous variants of the virus, and in the sense this may make evolutionarily advantageous some variants which are more lethal, and all in all reduce the vaccine coverage among senior citizens.

A more detailed discussion would require the study of \emph{dynamical models}; these however should make precise assumptions about several issues (time decay of vaccine protection, rate of appearance of new variants, probability that these are more infective or more lethal, etc.); the scope of the present paper is instead to point out some \emph{general} feature and mechanism which can be at play in the epidemic dynamics when both vaccination and evolution are taken into account; these are based on simple considerations.

Needless to say, the very simple considerations presented here do not suffice to reach any firm conclusion in this sense; but we believe they should stimulate further and more detailed work in this direction.

Finally, I stress that I have tried to keep the discussion to an extremely simple level, also in order to show that the mechanisms involved are very basic; the reader can find some guidance to the existing literature on this matter in Appendix \ref{app:previous}.

\subsection*{Acknowledgements}

This work was performed at SMRI, providing as usual relaxed working conditions. The author is also a member of GNFM-INdAM.
I thank M. Cadoni (who bears of course no responsibility for the opinions expressed here) for useful discussions and criticism.

I also thank an unknown Referee for his/her patience with the many mistakes in the first version of the paper and for suggesting further investigation into the relation with lethality, see Appendix \ref{sec:appleth}.

The author has been duly COVID-vaccinated as mandated by the Italian Ministry of University.

\begin{appendix}

\section{Some previous work}
\label{app:previous}

We are of course not the first to investigate the question discussed here, and more generally the interplay between vaccination and pathogen evolution, both in general terms and in relation to specific diseases. A short note is surely not the appropriate place to provide a review, but we would like to give a list of some of the relevant references and previous work on the same theme.

A general discussion of the interplay between pathogen evolution and epidemic dynamics was provided e.g. by the influential papers by Girvan, Callaway, Newman and Strogatz \gcite{GCNS} and by Gog and Grenfell \gcite{Gog1}; their line of thought was followed by many authors, in particular concerning Influenza A epidemics for which the interplay between evolution and competition between different strains is specially fascinating and relevant, see e.g. \gcite{Gog2,Meyers,Tria,Boni,Koelle,Bianconi,Karrer,Blake}; they were not concerned with vaccines.

The problem considered here -- and the investigations mentioned above and recalled in a somewhat greater detail in the lines below -- depend of course on the different protection offered by vaccines w.r.t. the natural protection provided by the immune system once the host has been in contact with the virus (provided the host survived, which is not always the case). The fact the former cannot be hoped to be at the same level of the latter -- also in terms of duration in time -- is well known and has even be dubbed (by Sabin himself) the ``Sabin's law  about live virus vaccines'' \gcite{Sabin0,Sabin1}. This was set down in the course of discussion about the possibility -- and the failures in the attempt -- to realize a HIV vaccine\footnote{As far as I know, this was the first context in which a mRNA approach was considered; I understand nowadays a new mRNA based HIV vaccine is under study, using the experience gained in the COVID vaccine formulation and production.}. See also \gcite{Nowak4} in this respect.

This however lies in the background of our discussion, and we have preferred not to really consider this aspect of the problem, but only the protection offered by the vaccine in itself against different variants, and the evolutionary effects of having a large fraction of the host population protected by a (necessarily imperfect) vaccine.

The possible evolutionary effects of wide-scale vaccination \gcite{GDevi} have been considered also for other diseases directly transmitted from human to human, as influenza \gcite{Boni,Gog} and avian influenza \gcite{Escorcia}; and also on diseases transmitted by an external carrier: this is e.g. the case of malaria \gcite{MGR}.

A problem which was discussed in detail is that of long-term effects of measles vaccine campaign, particularly in geographical situations where there is no hope of -- even local -- eradication in a reasonable amount of time (which means one or two decades, see \gcite{ML2} for a terse explanation of this timeline); see e.g. \gcite{MLA1,MLA2}.

In connection with the new problems raised by the COVID pandemic, which is definitely different from those faced in the past, one is however more interested in general theoretical discussions. A rich literature exist also for this type of discussions.

We mention here early works by the group around May and Nowak \gcite{Nee,MayNow,Nowak3,Nowak1,Nowak2} and by McLean \gcite{ML0,ML1} (see also the related, and already mentioned, works by McLean and Anderson \gcite{MLA1,MLA2}), together with some more recent ones by Gandon, Day and coworkers, see \gcite{GMNR,Proulx,Andre} and in particular \gcite{GTIF}.

All these works are over ten years old, and some of these are nearly thirty years old -- which shows that the problem is on the board since a substantial time. More recent discussions in general terms are provided e.g. by reff. \gcite{Read, Cressler, Parsons, Price}. See also \gcite{Bell,Miller} for contemporary, also COVID-related, discussions.

It should be noted, however, that the general discussions mentioned here cannot be directly applied to the COVID framework, due to the specific features of both the viral infection (a large number of asymptomatic ones, usually very light consequences on young people, etc.), and available vaccines (targeting the spike and not the in-cell active part of the virus). Moreover, many of these consider a ``mean field'' approach; the interplay between actual mass vaccination (as we are witnessing in several countries for COVID) and evolution relies, for a virus like COVID, on rare mutations and thus would require a more substantially probabilistic approach.

\section{Medical versus epidemiological protection}
\label{app:asprot}

As well known, the available COVID vaccines offer a rather good medical protection to vaccinees, i.e. these are shielded from developing serious consequences to a rate which differs for different vaccines (and decays in time \gcite{Naaber}) but is however over 90\%; on the other hand little is known about protection against asymptomatic infection, which is a key factor in epidemiological terms \gcite{Swan}.

Quite surprisingly, but maybe understandably given the large number of individuals involved, it seems all studies about vaccine effectiveness did not involve a full screening of the vaccine and the control groups.

In order to make more clear this point, we will quote a very recent and quite popular (also in the general press, thanks to popularization by Nature \gcite{natwan}) study \gcite{Brux} yielding reassuring results about the permanence in time of vaccine protection (in particular by one producer, which was also the funder of the study). In this study it is stated that (page 7 therein):

{\it ``For this interim analysis, asymptomatic COVID-19 cases were identified through positive SARS-CoV-2 tests ordered for individuals without COVID-19 symptoms (e.g., routine screening prior to procedures or hospital admission at KPSC, elective screening of KPSC employees, or testing requested for any other reason); these test orders were not used for individuals with symptoms. The remainder of the COVID-19 cases were considered symptomatic COVID-19 cases.''}

This shows that there was no systematic screening for asymptomatic infections. One should add that this study involved 352,878 vaccinated and 352,878 unvaccinated individuals, so that systematic screening looks unpractical; however, the same lack of systematic screening for asymptomatic infections appears to have occurred in the first tests, based on much smaller groups.

It should also be said that vaccinated people are most probably more prone than unvaccinated ones to accept situations in which they risk contagion (e.g. crowded transports), so real protection against asymptomatic infection could be higher than it appears from a purely statistical counting. See also the discussion in The Lancet \gcite{lancetdisc} about the correct way of statistically estimating this.

\section{Discussion in terms of lethality}
\label{sec:appleth}

Our discussion was conducted in terms of the \emph{fitness} $\phi$ of different strains, and we showed that when comparing the fitness of two strains, the strain with lower ``naked'' fitness (that is, fitness in a vaccine-naive population) -- but able to evade the vaccine action -- may happen to become the one with the higher one when the virus is confronted with an environment in which vaccinees abound. We also mentioned that this is worrying in that generally speaking there is an inverse relation between fitness and lethality, so that in such a case the more lethal strain would end up having an evolutionary advantage.

It may be argued that this discussion would be more to the point if we were able to give a relation between the  difference in lethality and the threshold for such an advantage of the more lethal strain to emerge.

It is clear that the exact nature of this relation will depend on the exact nature of the relation between fitness and lethality; which is just the point we have avoided to discuss, in order to show the point discussed in the main text depends only on general features (that is, the inverse relation between ``naked'' fitness and lethality on the one hand, and the lack of strict relation between ``naked'' fitness and ability to evade the vaccine action on the other).

One can however try to explore the simplest detailed model -- actually, the only one in which analytical considerations can be formulated with a rather direct discussion -- in order to gather some information on the possible consequences of the general mechanism depicted here in terms of possible lethality increase as different strains become dominant.

The simplest form of the relation between fitness (we stress once again this is ``naked'' fitness, i.e. without taking into account the vaccine effect) and lethality for different strains of a virus, satisfying the general criterion mentioned above, is a \emph{linear} one:
\beql{eq:pllin} \phi \ = \ \a \ \( 1 \ - \ \la \) \ . \eeq
This says that the maximal fitness for this family of viruses is $\a$, and that the fitness reduces to zero for a virus having lethality $\la = 1$, i.e. killing all of its hosts. A slightly modified form of this relation is
\beql{eq:plk} \phi \ = \ \a \ \( 1 \ - \ \la \)^k \ , \eeq with $k$ some positive parameter (see Fig.\ref{fig:philambda}).

Note that this is promptly inverted to give
\beq \la \ = \ 1 \ - \ \( \frac{\phi}{\a} \)^{1/k} \ . \eeq

In the following, we will slightly change our notation, and will write
$$ \psi_i \ := \ \phi_i (\chi ) \ , $$ so that $\phi_i := \phi_i (0) $ with no possible confusion.
This means that
\beql{eq:psiphi} \psi_i \ = \ \( 1 \ - \ \ga_i \ \chi \) \ \phi_i \ . \eeq
Moreover, we are interested in the case where
\beq \phi_2 \ < \ \phi_1 \ , \ \ \psi_2 \ > \ \psi_1 \ . \eeq
We assume the first of these relations is satisfied, and investigate about the second being satisfied or otherwise.
We will always write, as above, $p : \phi_2 / \psi_1 < 1$, $q = \ga_2 / \ga_1 < 1$. Thus the relation $\psi_2 > \psi_1$ reads
\beq p \ (1 \ - \ \chi \, q \, \ga_1) \ > \ (1 \ - \ \chi \ \ga_1 ) \ . \eeq

\begin{figure}
  \includegraphics[width=150pt]{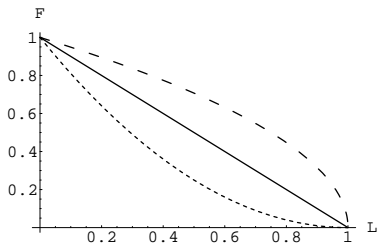}\\
  \caption{Relation between lethality (L) and fitness (F) according to \eqref{eq:plk}. We plot the curves for $k=1$ (solid), for $k= 2$ (dashed), and for $k = 1/2$ (dotted).}\label{fig:philambda}
\end{figure}

\subsection{The linear case}

Let us first consider the linear case, i.e. \eqref{eq:pllin}. We will moreover set
$$ r \ := \ \frac{\la_1}{\la_2} \ \in \ [0,1] \ . $$
In order to further simplify notation, we also write
$$ \ga_1 \ \equiv \ \ga \ , \ \ \ \la_2 \ \equiv \ \la \ . $$
Now the relation $\psi_2 > \psi_1$ reads
\beql{eq:lalin} (1 \ - \ \chi \, q \, \ga ) \ (1 \, - \, \la ) \ > \ (1 \, - \, \chi \, \ga) \, (1 \, - \, r \, \la) \ . \eeq
Thus the critical value for $\chi$ is in this case
\beql{eq:CS1} \chi_* \ = \ \frac{\la \ (1 - r)}{\ga \ [ (1-q) \, + \, \la \, (q-r) ]} \ . \eeq
We should recall that, for this to have actually the meaning of a critical vaccination rate
we should require
$$ 0 \le \chi_* \le 1 \ . $$
The condition $\chi_* \le 1$ amounts to $\la \le 1$, and is thus automatically satisfied; as for the condition $\chi_* > 0$, in view of $r <1$ it requires to have
$$ \la \ > \ \frac{1-q}{q - r} \ . $$

Plots for $\chi_*$ as given by \eqref{eq:CS1}, and considered as a function of $(r,q)$ for different given values of $(\ga,\la)$ are provided in Fig.\ref{fig:CS1cont}. These show that there is indeed the possibility that the more lethal variant becomes predominant if it is more effective in eluding the vaccine action and the vaccination rate in the population is high enough, and give an idea of the quantitative behavior if the vaccination threshold $\chi_*$.

\begin{figure}
  \includegraphics[width=150pt]{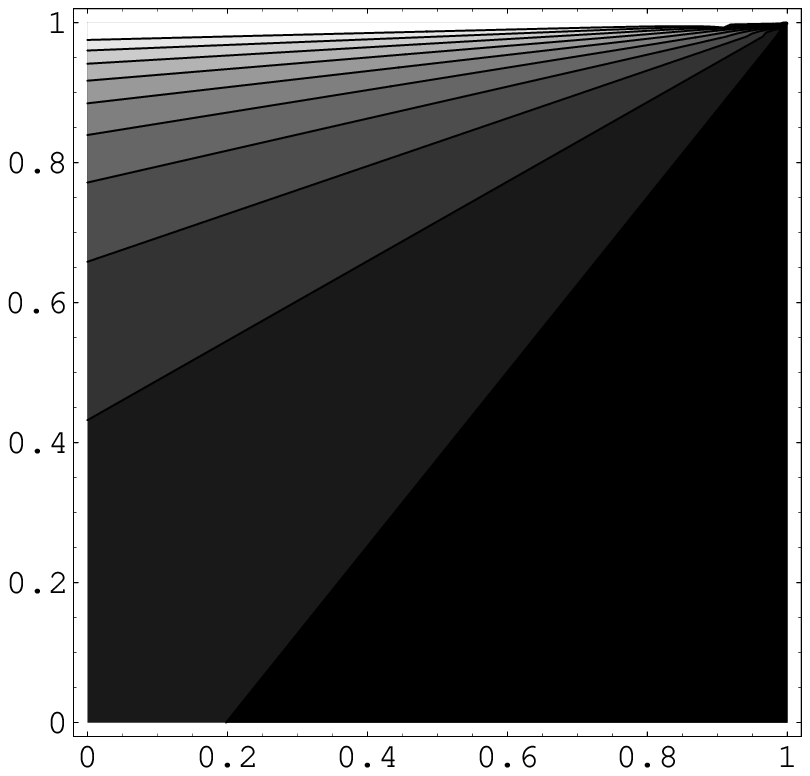}\\
  \includegraphics[width=150pt]{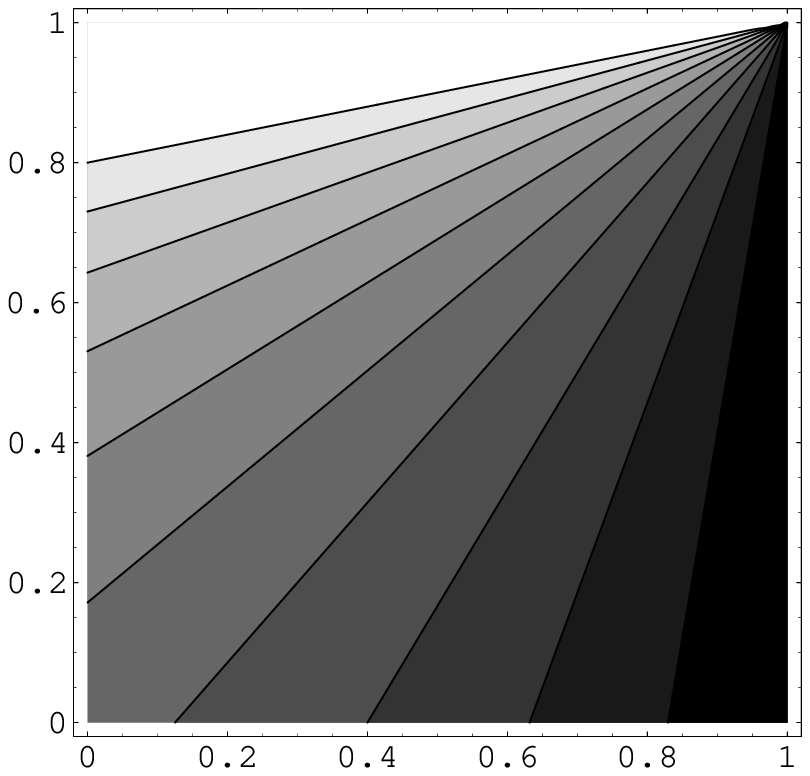}\\
  \includegraphics[width=150pt]{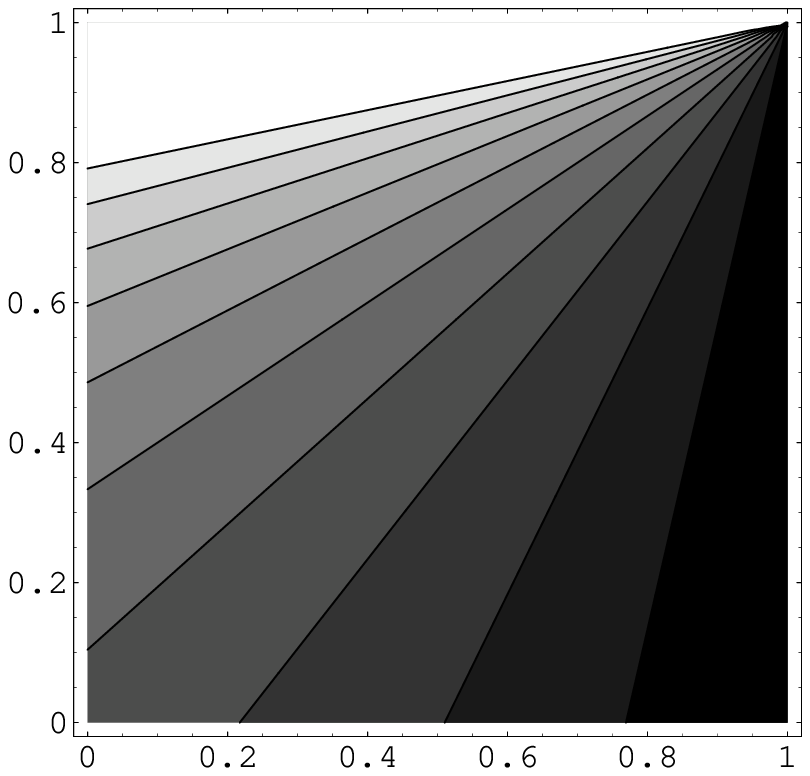}\\
  \caption{Contour plots of of $\chi_*$ defined in eq.\eqref{eq:CS1} for the linear case \eqref{eq:pllin} as a function of $r$ and $q$ for given $\ga$ and $\la$. We plot the cases $\ga=0.9$, $\la = 0.1$ (upper), $\ga = 0.75$, $\la = 0.3$ (middle), and $\ga = 0.5$, $\la = 0.2$ (lower).}\label{fig:CS1cont}
\end{figure}

\subsection{The nonlinear case}

A similar discussion can be conducted in the nonlinear case \eqref{eq:plk}. In these cases the critical value is given by
\beql{eq:chi*gen} \chi_* \ = \ \frac{( 1 - \la )^k \ - \ (1 - r \la)^k}{\ga [ (1 - \la )^k \, q \ + \ (1 -r \la)^k} \ . \eeq
We will just plot the resulting functions -- in the range $0 \le \chi_* \le 1 $ in the cases $k=2$ and $k=1/2$, respectively in Fig.\ref{fig:CS2} and in Fig.\ref{fig:CS3} (see also Fig.\ref{fig:philambda}).

The qualitative pictures which emerges is not substantially different from the one seen in the linear case. In particular, it appears that the emergence of the more lethal variant as dominant one due to better performance against vaccines when the vaccination rate in the population is sufficiently high is still possible -- and not an artifact of the linear model.

\begin{figure}[t]
  \includegraphics[width=150pt]{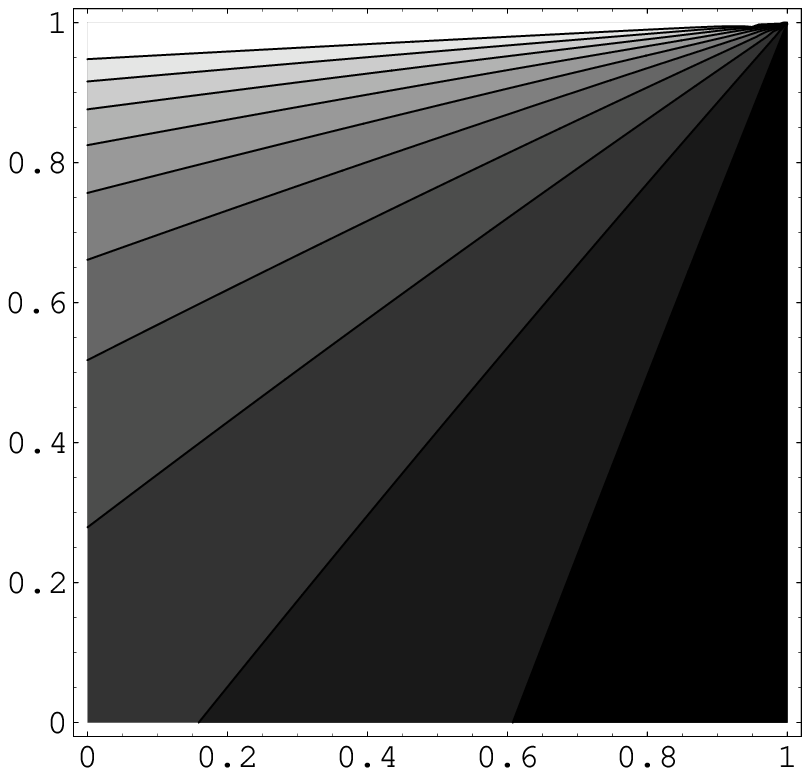}\\
  \includegraphics[width=150pt]{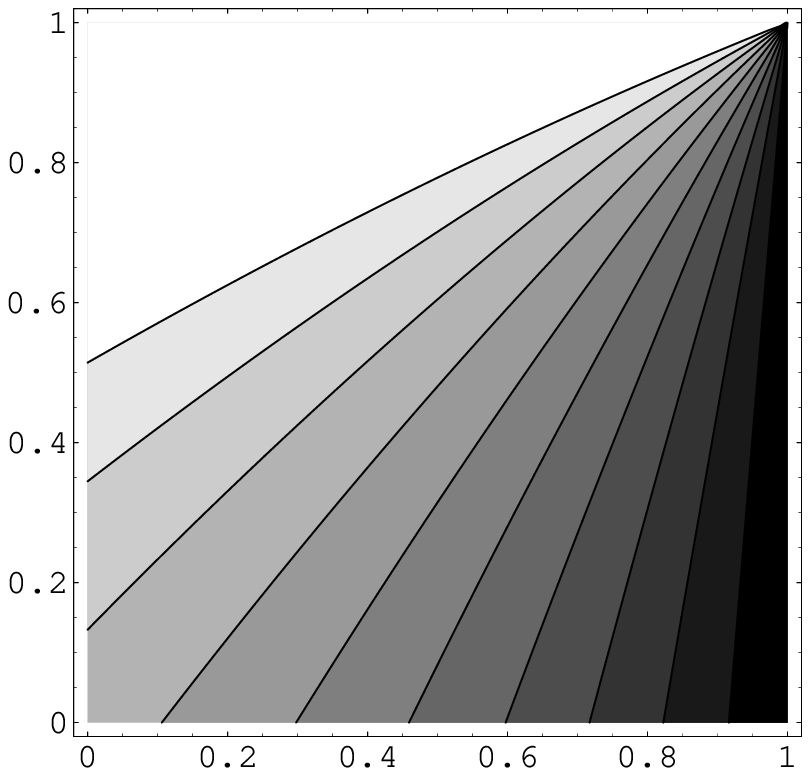}\\
  \includegraphics[width=150pt]{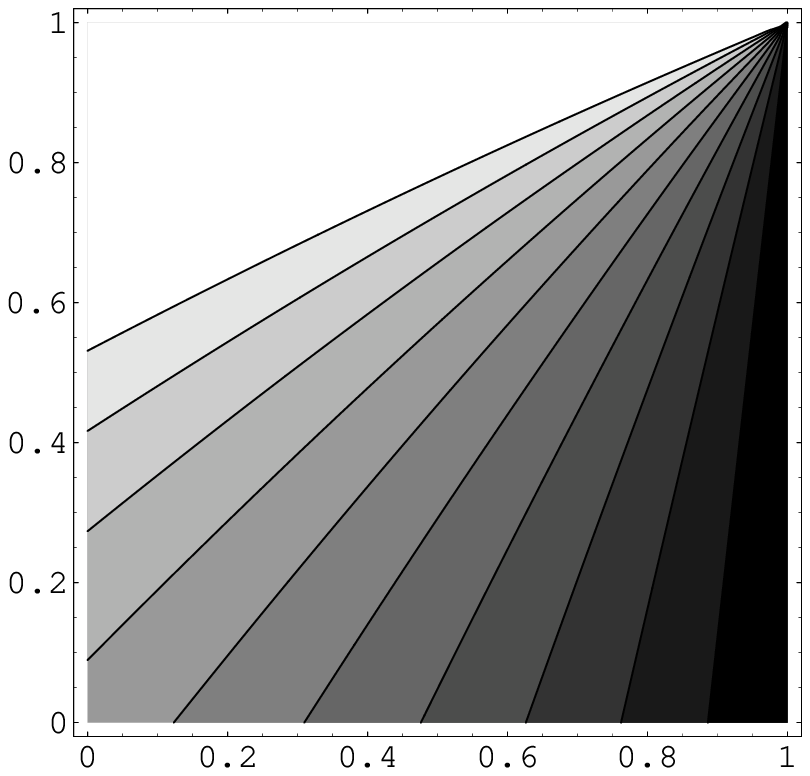}\\
  \caption{Contour plots of of $\chi_*$ defined in eq.\eqref{eq:CS1} for the nonlinear case \eqref{eq:plk} with $k=2$ as a function of $r$ and $q$ for given $\ga$ and $\la$. We plot the cases $\ga=0.9$, $\la = 0.1$ (upper), $\ga = 0.75$, $\la = 0.3$ (middle), and $\ga = 0.5$, $\la = 0.2$ (lower).}\label{fig:CS2}
\end{figure}

\begin{figure}[t]
  \includegraphics[width=150pt]{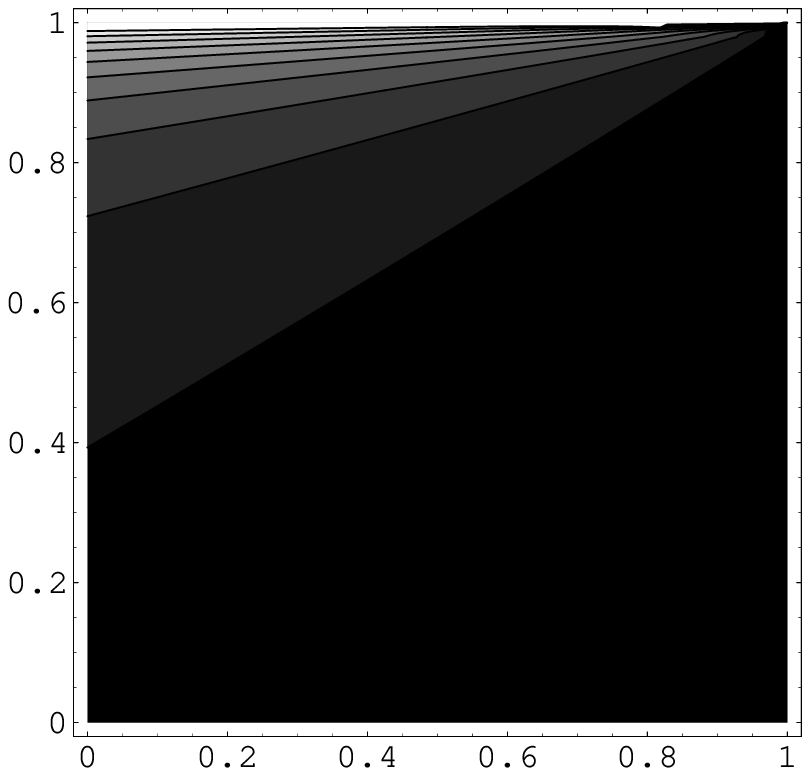}\\
  \includegraphics[width=150pt]{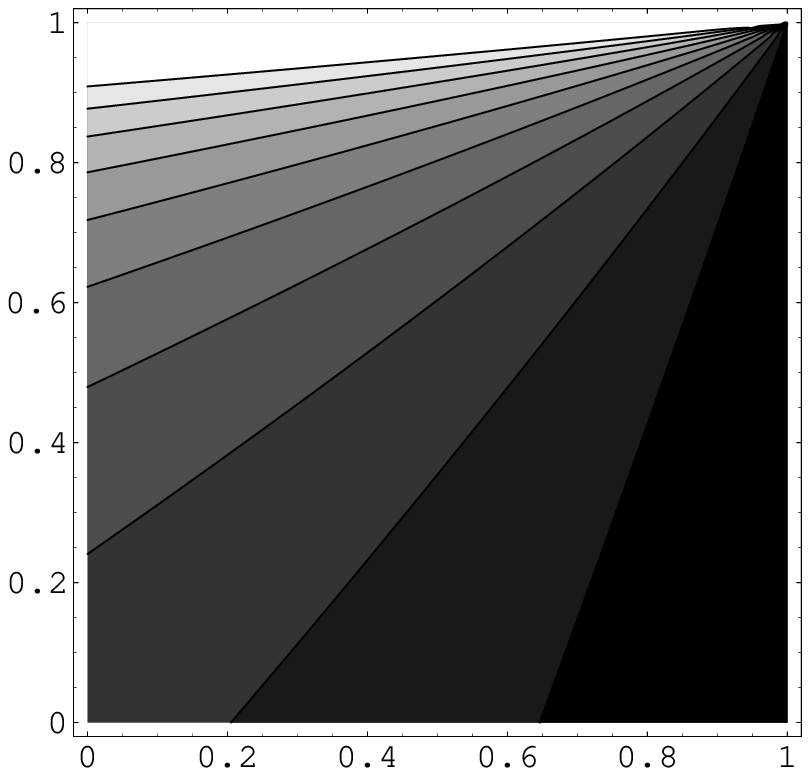}\\
  \includegraphics[width=150pt]{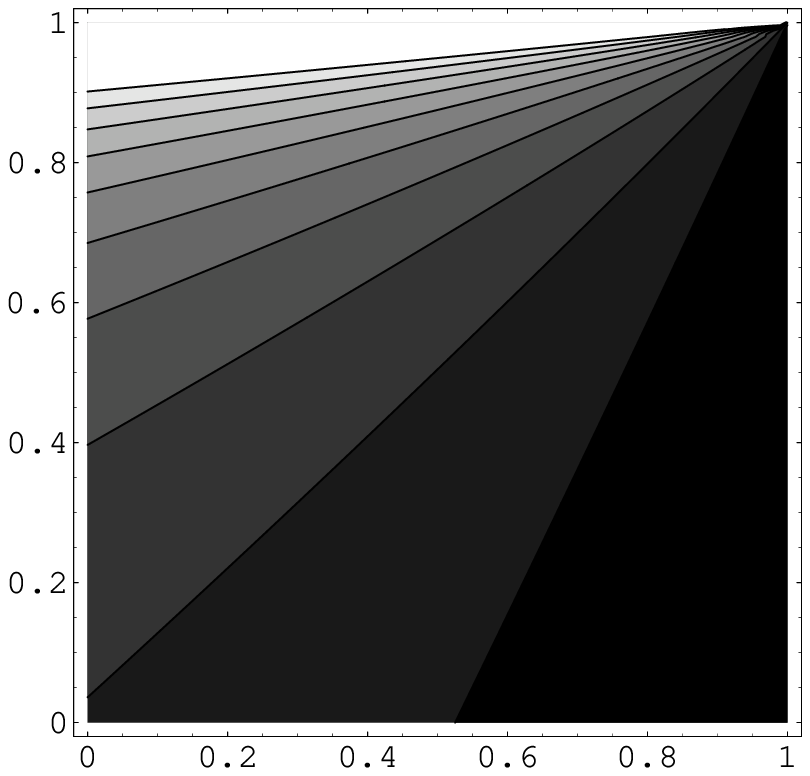}\\
  \caption{Contour plots of of $\chi_*$ defined in eq.\eqref{eq:CS1} for the nonlinear case \eqref{eq:plk} with $k=1/2$ as a function of $r$ and $q$ for given $\ga$ and $\la$. We plot the cases $\ga=0.9$, $\la = 0.1$ (upper), $\ga = 0.75$, $\la = 0.3$ (middle), and $\ga = 0.5$, $\la = 0.2$ (lower).}\label{fig:CS3}
\end{figure}

It should be stressed that -- luckily -- in many case one has $\la \simeq 0$. In the small $\la$ approximation, i.e. expanding \eqref{eq:chi*gen} as a Taylor series in $\la$ around $\la = 0$ and truncating this at second order, we get

\newpage

\begin{eqnarray*}
\chi_* &\approx& \frac{k}{\ga} \ \frac{1-r}{1-q} \ \la \\
& & \ + \ \frac{k}{\ga} \ \frac{(1-r) \, [k (1 + q) (1 - r) - (1 - q) (1 + r)]}{2 \ (1 - q)^2} \ \la^2 \ . \end{eqnarray*}

\end{appendix}

\end{document}